\documentclass[preprints,article,accept,moreauthors,pdftex]{Definitions/mdpi}

\usepackage{amsmath}
\usepackage{amsfonts}
\usepackage{amssymb}
\usepackage{graphicx}
\usepackage[normalem]{ulem}
\definecolor{myred1}{RGB}{255, 0, 0}
\definecolor{myyellow1}{RGB}{255, 255, 219}
\definecolor{mygreen1}{RGB}{0, 255, 0}
\definecolor{mygreen2}{RGB}{0, 126, 0}
\definecolor{myblue1}{RGB}{0, 0, 255}
\definecolor{mypink1}{RGB}{255, 105, 180}
\definecolor{mypurple1}{RGB}{128, 0, 128}
\definecolor{mygray1}{RGB}{110, 123, 139}
\usepackage{xcolor}

\firstpage{1}
\pubvolume{xx}
\issuenum{1}
\articlenumber{5}
\pubyear{2019}
\copyrightyear{2019}
\history{Received: date; Accepted: date; Published: date}

\pdfoutput=1

\Title{On the gauge fixing in the Hamiltonian analysis of general teleparallel theories}

\Author{Daniel Blixt $^{1,*}$\orcidA{}, Manuel Hohmann $^{1,\dagger}$ and Christian Pfeifer $^{1,\ddagger}$}

\AuthorNames{Firstname Lastname, Firstname Lastname and Firstname Lastname}

\address[1]{%
$^{1}$ \quad Laboratory of Theoretical Physics, Institute of Physics, University of Tartu, W. Ostwaldi 1, 50411 Tartu, Estonia}

\corres{Correspondence: blixt@ut.ee}

\firstnote{manuel.hohmann@ut.ee}
\secondnote{christian.pfeifer@ut.ee}

\abstract{The covariant formulation of teleparallel gravity theories must include the spin connection, which has 6 degrees of freedom. One can, however, always choose a gauge such that the spin connection is put to zero. In principle this gauge may affect counting of degrees of freedom in the Hamiltonian analysis. We show for general teleparallel theories of gravity, that fixing the gauge such that the spin connection vanishes in fact does not affect the counting of degrees of freedom. This manifests in the fact that the momenta of the Lorentz transformations which generate the spin connection are fully determined by the momenta of the tetrads.}


\begin{document}


\section{Introduction}
General relativity (GR) has successfully passed a huge amount of experimental tests, which probe the nature of gravity, up to today. Despite this success there are still many open questions associated with our understanding of gravity. Firstly, general relativity is highly non-renormalizable, so it can not be formulated as a quantum field theory in the same way as it is done for the other fundamental forces, and thus can not directly be embedded into the standard model of particle physics. Secondly, there is strong evidence for inflation. To describe this one is led to either introduce an extra field (like the inflaton) in the early universe or modify the laws of gravity. The latter gives a better fit to the data \cite{Akrami:2018odb}. Thirdly, there are tensions in cosmological data such as the value of the Hubble constant \cite{Aghanim:2018eyx,Riess:2018uxu}, which needs to be explained. Furthermore, the standard model of cosmology is based on the $\Lambda$CDM model, whose main ingredients are cold dark matter particles and a cosmological constant as dark energy, to explain the dark sector of our universe. However, also this model faces some issues, where the biggest issue probably is the smallness of the cosmological constant.

In order to deal with the aforementioned issues modified theories of gravity have been studied. Most are based on the formulation of general relativity in terms of the Levi-Civita connection, which is induced by a spacetime metric. However, general relativity has other equivalent formulations, based on connections that are not induced by the metric. One of these is called ``symmetric teleparallel equivalent of general relativity'' (STEGR) and uses a flat (no curvature) and torsion free connection with non-metricity ($\nabla g_{\mu\nu}\neq 0$). Another is called ``teleparallel equivalent of general relativity'' (TEGR) and employs a flat metric compatible connection with torsion. The Lagrangian of STEGR is given by the so-called non-metricity scalar $Q$, while the Lagrangian of TEGR by the so-called torsion scalar $T$. These reformulations of Einstein's theory of general relativity are sometimes referred as "the geometrical trinity" \cite{BeltranJimenez:2019tjy}.

Due to the experimental success of general relativity we need to formulate modified theories of gravity such that they are compatible with experimental tests on solar system scales. That is, they should not deviate too much from general relativity on these scales. Since general relativity can equivalently be formulated in different geometries, we have the freedom to choose which geometry we want to formulate modified theories of gravity in. After modifying general relativity, the modified theories will in general be in-equivalent.

For example, popular modifications of general relativity are to consider functions of the defining Lagrangian. In the three different formulations this amounts to consider as Lagrangian either $f(R)$, where $R$ is the Ricci scalar of the Levi-Civita connection, $f(T)$ or $f(Q)$, which lead to non-equivalent theories. The reason for this is that they differ by a boundary term, which can no longer be completely neglected when a function is acting on the original GR, STEGR or TEGR Lagrangian.

In this work we will consider the Hamiltonian analysis of modified theories of gravity in the teleparallel framework. The Hamiltonian analysis gives the number of degrees of freedoms in a theory. However, in the so-called $f(T)$ theories of gravity disputing results have been found for this number. Where it was claimed in \cite{Chen:2014qtl,Li:2011rn} that the theory has 5 degrees of freedom. More recent work, on the contrary, found that $f(T)$ has 3 degrees of freedom \cite{Ferraro:2018tpu}. The aforementioned works were, however, done in a gauge where the spin connection is put to zero, which is not the covariant formulation of teleparallel gravity \cite{Krssak:2015oua,Golovnev:2017dox}. We show in this work, for general covariant  teleparallel theories, that the spin connection momenta are determined by the tetrad momenta .\\
\indent In Section \ref{sec:GeneralTheory} we display the most general teleparallel gravity theories we consider in this article. Section \ref{sec:ConjMomentum} is devoted to derive the conjugate momenta, and to show that the gauge fixing does not affect the counting of numbers of degrees of freedom. A concrete example is provided in Section \ref{sec:NGR} with an explicit expression for the Hamiltonian. Finally, discussion and concluding remarks are made in Section \ref{sec:Discussion}.\\
\indent We use the following conventions. Greek indices $\mu,\nu,\rho...$ denotes global coordinate indices which are raised and lowered with the metric $g_{\mu\nu}$, capital Latin indices denotes Lorentz indices raised and lowered with the Minkowski metric $\eta_{AB}$, and small Latin indices are spatial indices and $0$ denotes the temporal index. The Minkowski metric $\eta_{AB}$ is taken to be $\mathrm{diag}(-1,1,1,1)$. Brackets $[]$ denote dependence on the explicit variables and their derivatives.

\section{Generalized theories of teleparallel gravity}
\label{sec:GeneralTheory}
The fundamental variables for teleparallel gravity theories are the tetrads (or vierbeins) $\theta^A$, and for the covariant formulation a curvature-free spin-connection $\omega^A{}_B$ is needed \cite{Krssak:2015oua,Golovnev:2017dox}. In local coordinates these variables can be expressed as
\begin{align}
	\begin{split}
	\theta^A&=\theta^A{}_\mu \mathrm{d}x^\mu,  \quad e_A=e_A{}^\mu \partial_\mu, \\ \omega^A{}_B&=\omega^A{}_{B\mu}\left[\Lambda^C{}_D\right]\mathrm{d}x^\mu=\Lambda^A{}_C\mathrm{d}\left(\Lambda^{-1}\right)^C{}_B=\Lambda^A{}_C\partial_\mu\left(\Lambda^{-1}\right)^C{}_B\mathrm{d}x^\mu,
	\end{split}
\end{align}
where $\Lambda^A{}_B$ are Lorentz matrices. Any Lorentzian metric can be expressed in terms of tetrads by the following relations
\begin{align}
	\begin{split}
	g_{\mu\nu}=g_{\mu\nu}\left(\theta^A{}_\mu\right)=\eta_{AB}\theta^A{}_\mu\theta^B{}_\nu, \quad g^{\mu\nu}=\eta^{AB}e_A{}^\mu e_B{}^\nu.
	\end{split}
\end{align}
The torsion components expressed in tetrad fields and the spin connection are
\begin{align}
\label{Torsion}
	T^{\rho}{}_{\mu\nu}=e_A{}^{\rho}T^A{}_{\mu\nu}[\theta^A{}_\mu,\mathrm{d}\theta^A{}_\mu,\Lambda^A{}_B,\mathrm{d}\Lambda^A{}_B]=e_A{}^\rho \left(\partial_{[\mu} \theta^A{}_{\nu]}+\omega^A{}_{B[\mu}\theta^B{}_{\nu]}\right)
\end{align}

 We can write a generic action made from the Torsion components $T^\rho{}_{\mu\nu}$ and the metric (which depend on the tetrad fields) as
\begin{align}
	\label{action}
	S\left[\theta^A{}_\mu,\Lambda_C{}^D\right]=\int \mathrm{d}^4 xL\left[\theta^A{}_\mu,\Lambda_C{}^D\right]=\int \mathrm{d}^4 x \ |\theta|f(g_{\sigma\tau},T^{\rho}{}_{\mu\nu}),
\end{align}
where $|\theta|:=\det (\theta^A{}_\mu)$ which is the normal volume element ($\sqrt{-g}$ in metric formalism).
This is the most general teleparallel gravity theory in 4 dimensions without introducing extra fields, and without breaking local Lorentz invariance, with all derivatives being of first order and coming from the torsion components, and includes the theories discussed in \cite{Bahamonde:2017wwk}. The analysis can easily be extended to higher dimensions.
In order to derive the conjugate momenta and make a canonical Legendre transformation to the Hamiltonian, we make use of the 3+1-decomposition analogous to \cite{Blixt:2018znp}. In this decomposition we have
\begin{align}
	g_{\mu\nu}=\begin{bmatrix}
	-\alpha^2+ \beta^i\beta^jh_{ij} & \beta_i\\
	\beta_j & h_{ij}
	\end{bmatrix}, \quad g^{\mu\nu}=\begin{bmatrix}
	-\frac{1}{\alpha^2} & \frac{\beta^i}{\alpha^2}\\
	\frac{\beta^j}{\alpha^2} & h^{ij}-\frac{\beta^i\beta^j}{\alpha^2}
	\end{bmatrix}.
\end{align}
The indices $i,j,...$ are spatial and run from 1 to 3 and are raised and lowered with the induced metric $h_{ij}$, i.e. $\beta_i=\beta^jh_{ij}$. For the tetrad fields (which are canonical variables for teleparallel gravity theories) we have
\begin{align}
	\theta^A{}_0 =\alpha \xi^A+\beta^i\theta^A{}_i,
\end{align}
where $\xi^A$ are components of the normal vector $n$ to the $x^0=\textrm{const}$ hypersurfaces in the dual tetrad basis \cite{Okolow:2011nq}
\begin{align}
	n=\xi^Ae_A, \quad \xi^A=-\frac{1}{6}\epsilon^A{}_{BCD}\theta^B{}_i\theta^C{}_j\theta^D{}_k\epsilon^{ijk}.
\end{align}
The components $\xi^A$ further satisfy
\begin{align}
	\eta_{AB}\xi^A\xi^B=\xi^A\xi_A=-1, \quad \eta_{AB}\xi^A\theta^B{}_i=\xi_A\theta^A{}_i=0.
\end{align}
Furthermore, the dual tetrads and the induced metric can be expressed as
\begin{align}
	e_A{}^0=-\frac{1}{\alpha}\xi_A, \quad e_A{}^i=\theta_A{}^i+\xi_A\frac{\beta^i}{\alpha}, \quad h_{ij}=\eta_{AB}\theta^A{}_i\theta^B{}_j.
\end{align}
For readability we sometimes suppress metrics which raises or lowers indices, even when indices are at non-canonical positions. For example $\theta_A{}^i=\eta_{AB}h^{ij}\theta^B{}_j\neq e_A{}^i=\theta_A{}^i+\xi_A\frac{\beta^i}{\alpha}$.

\section{Conjugate momenta}
\label{sec:ConjMomentum}
To derive the conjugate momenta we note that time derivatives always appear in $T^{\rho}{}_{0i}=-T^\rho{}_{i0}=e_A{}^\rho T^A{}_{0i}$ due to the antisymmetric property of the torsion components $T^\rho{}_{00}=0$.  Time derivatives act on tetrad fields $\theta^A{}_i$ and Lorentz matrices $\Lambda^A{}_B$ and explicitly it reads
\begin{align}
\label{velocitydep}
	T^A{}_{0i}=\partial_0 \theta^A{}_i+\Lambda^A{}_{C}\partial_0\left(\Lambda^{-1}\right)^C{}_B\theta^B{}_i-\partial_i \theta^A{}_0-\Lambda^A{}_{C}\partial_i\left(\Lambda^{-1}\right)^C{}_B\theta^B{}_0.
\end{align}
One immediately finds that time derivatives never act on temporal tetrads ($\theta^A{}_0$) nor lapse and shifts ($\alpha,\beta$). They only act on the spatial tetrads $\theta^A{}_i$ and Lorentz matrices $\Lambda^A{}_B$. Hence, the conjugate momenta only need to be defined for these variables. The conjugate momenta with respect to the spatial tetrad fields are defined by
\begin{align}
	\label{ConjMom}
	\pi_A{}^i:=\frac{\partial L}{\partial \partial_0 \theta^A{}_i}=|\theta| \frac{\partial f}{\partial T^\mu{}_{0j}}\frac{\partial T^\mu{}_{0j}}{\partial \partial_0 \theta^A{}_i}=|\theta| e_A{}^\mu\frac{\partial f}{\partial T^\mu{}_{0i}}.
\end{align}
Since the Lorentz matrices only have 6 independent components, we introduce an auxiliary antisymmetric field which preserves the Lorentz symmetries and thus also those of the spin connection
\begin{align}
	a_{AB}:=\eta_{AC}\omega^C{}_{B0}=\eta_{C[A}\Lambda^C{}_{|D|}\partial_0 \left(\Lambda^{-1}\right)^D{}_{B]} \Leftrightarrow \partial_0 \Lambda^A{}_B=a_{CD}\eta^{A[D}\Lambda^{C]}{}_B.
\end{align}
The conjugate momenta of the independent components of the Lorentz matrices are hence represented by
\begin{align}
\label{momentarelation}
	\hat{\pi}^{AB}:=\frac{\partial L}{\partial a_{AB}}=|\theta|\frac{\partial f}{\partial T^\mu{}_{0i}}\frac{\partial T^\mu{}_{0i}}{\partial a_{AB}}=-\pi_C{}^i\eta^{C[B}\theta^{A]}{}_i.
\end{align}
This can be realized from
\begin{align}
	\begin{split}
			\frac{\partial L}{\partial a_{AB}}&=\frac{\partial L}{\partial \partial_0 \Lambda^C{}_D}\frac{\partial \partial_0 \Lambda^C{}_D}{\partial a_{AB}}=\frac{\partial L}{\partial T^\mu{}_{0i}}\frac{\partial T^\mu{}_{0i}}{\partial \partial_0 \Lambda^C{}_D}\frac{\partial \partial_0 \Lambda^C{}_D}{\partial a_{AB}}\\
			&=-\frac{\partial L}{\partial T^\mu{}_{0i}}\frac{\partial T^\mu{}_{0i}}{\partial \partial_0 \theta^C{}_j}\left[\theta^D{}_j \left(\Lambda^{-1}\right)^F{}_D \right]\frac{\partial \partial_0 \Lambda^C{}_F}{\partial a_{AB}}\\
			&=-|\theta|\frac{\partial f}{ \partial T^\mu{}_{0i}}e_C{}^\mu{}\left[\theta^D{}_i \left(\Lambda^{-1}\right)^F{}_D\right]\eta^{C[B}\Lambda^{A]}{}_F.
	\end{split}
\end{align}
The conjugate momenta $\pi^A{}_i$ and $\hat{\pi}^{AB}$ are hence manifestly algebraically related to each other. This means that we need to add equation \eqref{momentarelation} as a Lagrange multiplier. Furthermore, it can be cumbersome to express the velocities into their conjugate momenta, but for new general relativity it has been shown how this can be done \cite{Blixt:2018znp}. To simplify we perform a transformation in which the spin connection vanishes and show that this transformation in this gauge is consistent with the constraints in the covariant formulation. This transformation is done by introducing new field variables $(\tilde{\alpha},  \tilde{\beta}^i, \tilde{\theta}^A{}_i, \tilde{\Lambda}^A{}_B)$ so that $\tilde{\theta}^A{}_i=\theta^B{}_i\left(\Lambda^{-1}\right)^A{}_B$, $\tilde{\alpha}=\alpha$, $\tilde{\beta}=\beta$, and $\tilde{\Lambda}^{A}{}_B=\Lambda^A{}_B$. It follows that $\tilde{a}_{AB}=a_{AB}$,  $\tilde{g}_{\mu\nu}=g_{\mu\nu}$, $\tilde{|\theta|}=|\theta|$ and that $\tilde{T}^ \rho{}_{\mu\nu}=\tilde{e}_A{}^\rho\partial_{[\mu}\tilde{\theta}^A{}_{\nu]} $. Furthermore,
\begin{align}
		\begin{split}
		\label{Lagrangiangauge}
				\tilde{L}&=\tilde{|\theta|}\tilde{f}(g_{\sigma\tau},T^\rho{}_{\mu\nu})=\tilde{|\theta|}f(\tilde{g}_{\sigma\tau},\tilde{T}^\rho{}_{\mu\nu})=|\theta|f(g_{\sigma\tau},\tilde{T}^\rho{}_{\mu\nu}),
		\end{split}
\end{align}
which manifestly is independent of the Lorentz matrices $\Lambda_A{}^B$. From this transformation we find that the conjugate momenta transforms as
\begin{align}
\label{transformedmomenta}
	\begin{split}
			\tilde{\pi}_A{}^i&=\frac{\partial \tilde{L}}{\partial \partial_0 \tilde{\theta}^{A}{}_i}=\pi_B{}^i\Lambda^B{}_A,\\
			\hat{\tilde{\pi}}^{AB}&=\frac{\partial \tilde{L}}{\partial a_{AB}}=\pi_C{}^i\eta^{C[B}\theta^{A]}{}_i+\hat{\pi}^{AB}.
	\end{split}
\end{align}
Inverting these formulas gives
\begin{align}
	\begin{split}
		\pi_A{}^i&=\tilde{\pi}_B{}^i\left(\Lambda^{-1}\right)^B{}_A,\\
		\hat{\pi}^{AB}&=\hat{\tilde{\pi}}^{AB}-\tilde{\pi}_D{}^i\left(\Lambda^{-1}\right)^D{}_C\eta^{C[B}\Lambda^{A]}{}_E\tilde{\theta}^E{}_i.
	\end{split}
\end{align}
Applying equation \eqref{momentarelation} to \eqref{transformedmomenta} shows that $\hat{\tilde{\pi}}^{AB}=0$ in the Weitzenböck gauge, and they are hence pure gauge degrees of freedom as expected from \cite{Krssak:2015oua,Golovnev:2017dox}. A vital point is now to show that the gauge fixing is imposed consistently with the constraints.
Hence, we need to show that $\{\hat{\tilde{\pi}}^{AB},\tilde{H}\}\approx 0$. The transformed Hamiltonian is defined as
\begin{align}
	\begin{split}
		\tilde{H}&=\tilde{\pi}_A{}^i\partial_0 \tilde{\theta}^A{}_i+\hat{\tilde{\pi}}^{AB}\tilde{a}_{AB}+{}^{\hat{\tilde{\pi}}}\lambda_{AB}\hat{\tilde{\pi}}^{AB}-\tilde{L}+\mathrm{primary \ constraints},
	\end{split}
\end{align}
where primary constraints need to be added (which differ from different theories).
Looking at the transformation behaviors of each term it is hence clear that $\{\hat{\tilde{\pi}}^{AB},H\}\approx 0$. The gauge fixing is hence consistent with the constraints and can not in any way affect the counting of degrees of freedom for teleparallel gravity theories.

\section{New general relativity}
\label{sec:NGR}
One interesting class of teleparallel gravity theories is the so-called "new general relativity" theory introduced in \cite{Hayashi:1979qx}. In this section we derive the Hamiltonian for "new general relativity" as was done in \cite{Blixt:2018znp}. In this section we work in the Weitzenböck gauge motivated by the preceding sections. Furthermore, we drop all $\ \tilde{} \ $ for readability. Assume that we want a teleparallel theory defined by equation \eqref{action} and only consider terms quadratic in the torsion components $T^\rho{}_{\mu\nu}$ without introducing parity violating terms. Then the action looks like
\begin{align}
	\begin{split}
		\label{NGRAction}
		S_{\mathrm{NGR}}=\int \mathrm{d}^4x|\theta|\left(c_1 T^\rho{}_{\mu\nu}T_\rho{}^{\mu\nu}+c_2T^{\rho}{}_{\mu\nu}T^{\nu\mu}{}_\rho+c_3T^\rho{}_{\mu\rho}T^{\sigma\mu}{}_\sigma\right).
	\end{split}
\end{align}
After a 3+1 decomposition it is found that
\begin{align}
	\begin{split}
		L_{\mathrm{NGR}}&=\frac{\sqrt{h}}{2\alpha}T^A{}_{i0}T^B{}_{j0}M^{i \ j}_{\ A \ B}+\frac{\sqrt{h}}{\alpha} T^A{}_{i0}T^B{}_{kl}\left[M^{i \ l}_{\ A \ B}\beta^k+2\alpha h^{il}\left(c_2\xi_B\theta_A{}^k+c_3\xi_A\theta_B{}^k\right)\right]\\
		&+\frac{\sqrt{h}}{\alpha}T^A{}_{ij}T^B{}_{kl}\beta^i \left[\frac{1}{2}M^{j \ l}_{\ A \ B}\beta^k+2\alpha h^{jl}\left(c_2\xi_B\theta_A{}^k+c_3\xi_A \theta_B{}^k\right)\right]+\alpha \sqrt{h}\ {}^3\mathbb{T}.
	\end{split}
\end{align}
Here
\begin{align}
	M^{i \ j}_{\ A \ B}:= -2(2c_1h^{ij}\eta_{AB}-(c_2+c_3)\xi_A\xi_B h^{ij}+c_2\theta_A{}^j\theta_B{}^i+c_3\theta_A{}^i\theta_B{}^j),
\end{align}
and
\begin{align}
	\begin{split}
		{}^3\mathbb{T}:=c_1\eta_{AB}T^A{}_{ij}T^B{}_{kl}h^{ik}h^{jl}+c_2\theta_A{}^i\theta_B{}^jT^A{}_{jk}T^B{}_{il}h^{kl}+c_3\theta_A{}^i\theta_B{}^j h^{kl}T^A{}_{ik}T^B{}_{jl}.
	\end{split}
\end{align}
The theory is covariant and the spatial derivatives can all be replaced (simultaneously) by the Levi-Civita covariant derivative $D_i$ associated with the induced metric such that $D_i h_{jk}=0$. Derivatives on the temporal parts of the tetrads ($\theta^A{}_0$) generally do not appear and hence the conjugate momenta for new general relativity are
\begin{align}
	\begin{split}
	\label{conjmomNGR}
	\frac{\alpha}{\sqrt{h}}\pi_A{}^i=\frac{\alpha}{\sqrt{h}}\frac{L_{\mathrm{NGR}}}{\partial \partial_0 \theta^A{}_i}=T^B{}_{0j}M^{i \ j}_{\ A \ B}+T^B{}_{kl}\left[M^{i \ k}_{\ A \ B}\beta^l+2\alpha h^{ik}\left(c_2 \xi_B \theta_A{}^l+c_3\xi_A \theta_B{}^l\right)\right].
	\end{split}
\end{align}
We can now define
\begin{align}
	\begin{split}
		S_A{}^i=\frac{\alpha}{\sqrt{h}}\pi_A{}^i+\left[D_k\left(\alpha \xi^B+\beta^m\theta^B{}_m\right)-T^B{}_{kl}\beta^l\right]M^{i \ k}_{ \ A \ B}-2\alpha T^B{}_{kl}h^{ik}\left(c_2 \xi_B \theta_A{}^l+c_3\xi_A \theta_B{}^l\right),
	\end{split}
\end{align}
so that $S_A{}^i$ is independent of velocities and equation \eqref{conjmomNGR} can equivalently be written as
\begin{align}
	\label{invmomentaNGR}
	S_A{}^i=\partial_0\theta^B{}_j M^{i \ j}_{\ A \ B}.
\end{align}
The remaining task is then to invert the $M^{i \ j}_{ \  A \ B}$ and solve for $\partial_0 \theta^B{}_j$. This is a rather non-trivial task, and hence, we refer to \cite{Blixt:2018znp} for details. Here we simply write out the possible primary constraints and the expression for the Hamiltonian. Existence, or non-existence of primary constraints depend on the specific values of $c_1,c_2,c_3$, related to the irreducible components under the rotation group into vectorial, antisymmetric, symmetric (but trace-free), and trace parts ($\mathcal{V},\mathcal{A},\mathcal{S},\mathcal{T}$). We define
\begin{align}
	A_{\mathcal{V}}=2c_1+c_2+c_3, \quad A_\mathcal{A}=2c_1-c_2, \quad A_\mathcal{S}=2c_1+c_2, \quad A_\mathcal{T}=2c_1+c_2+3c_3,
\end{align}
and an index $\mathcal{I}=\mathcal{V},\mathcal{A},\mathcal{S},\mathcal{T}$. Putting any of the $A_\mathcal{I}=0$ gives rise to primary constraints.
\begin{align}
	A_\mathcal{V}=0 \implies {}^\mathcal{V}C^i&:=S_A{}^i\xi^A=0,\\
	A_\mathcal{A}=0 \implies {}^\mathcal{A}C^{ij}&:=S_A{}^k\theta^A{}_{[j}h_{i]k}=0,\\
	A_\mathcal{S}=0\implies {}^\mathcal{S}C^{ij}&:=S_A{}^k\theta^A{}_{(j}h_{i)k}-\frac{1}{3}S_A{}^k\theta^A{}_kh_{ij}=0,\\
	A_\mathcal{T}=0\implies {}^\mathcal{T}C&:=S_A{}^i\theta^A{}_i=0.
\end{align}
The important thing to note is that if any of these primary constraints are imposed, they need to be added as Lagrange multipliers in the Hamiltonian. For new general relativity, the expression for the Hamiltonian is
\begin{align}
	\begin{split}
		H&=\alpha \sqrt{h}\left(B_\mathcal{V}\frac{{}^\mathcal{V}C_i{}^\mathcal{V}C^i}{4}-B_\mathcal{A}\frac{{}^\mathcal{A}C_{ij}{}^\mathcal{A}C^{ij}}{4}-B_\mathcal{S}\frac{{}^\mathcal{S}C_{ij}{}^\mathcal{S}C^{ij}}{4}-B_\mathcal{T}\frac{3{}^\mathcal{T}C{}^\mathcal{T}C}{4}-{}^3\mathbb{T}-\frac{\xi^AD_i\pi_A{}^i}{\sqrt{h}}\right)\\
		&-\beta^k\left(T^A{}_{jk}\pi_A{}^j+\theta^A{}_kD_i\pi_A{}^i\right)+D_i\left[\pi_A{}^i\left(\alpha\xi^A+\beta^j\theta^A{}_j\right)\right],
	\end{split}
\end{align}
where
\begin{align}
	\begin{split}
	 	B_{\mathcal{I}}=\begin{cases}
	 	\frac{1}{A_\mathcal{I}} \quad \mathrm{if} \ A_\mathcal{I} \neq 0\\
	 	0 \quad \ \ \  \mathrm{if} \ A_\mathcal{I}=0.
	 	\end{cases}
	\end{split}
\end{align}
This is, however, not the final Hamiltonian. As mentioned before, Lagrange multipliers related to primary constraints need to be added. Furthermore, the analysis might further provide secondary, tertiary, \ldots constraints after the evaluation of the Poisson brackets. This also needs to be added.
\section{Discussion}
\label{sec:Discussion}
We showed that for a very general class of teleparallel gravity theories one is allowed to fix the gauge such that the spin connection vanishes without affecting the counting of degrees of freedom in the theory. This significantly simplifies the Hamiltonian analysis of teleparallel gravity theories, assuring that the result does not differ from the covariant formulation. Furthermore, this justifies previous work where this gauge choice has been implemented in the analysis. Since the Hamiltonian analysis tends to be very cumbersome, it is highly suggestive to use this result and put the spin connection to zero in theories covered by this analysis. If one looks at more general teleparallel gravity theories (for example addition of extra fields or more dimensions, as they are discussed in the literature \cite{Hohmann:2018vle,Hohmann:2018dqh,Hohmann:2018ijr}) one can follow the same approach in order to figure out if the gauge fixing affects the counting of degrees of freedom.

\vspace{6pt}



\authorcontributions{Writing-original draft preparation: D.B.; conceptualization, methodology, validation, formal analysis, investigation, writing-review and editing: D.B., M.H., and C.P.; supervision: M.H., and C.P.; project administration, funding acquisition: M.H. }

\funding{This work was supported by the Estonian Research Council through the Personal Research Funding project PRG356 and by the European Regional Development Fund through the Center of Excellence TK133 ``The Dark Side of the Universe''.}

\acknowledgments{The authors would like to thank María José Guzmán and Martin Krššak for fruitful discussions.}

\conflictsofinterest{The authors declare no conflict of interest.}

\abbreviations{The following abbreviations are used in this manuscript:\\

\noindent
\begin{tabular}{@{}ll}
GR & General relativity\\
STEGR & Symmetric teleparallel equivalent of general relativity\\
TEGR & Teleparallel equivalent of general relativity
\end{tabular}}

\appendixtitles{no} 


\reftitle{References}





\end{document}